# Momentum-Transfer Model of Valence-Band Photoelectron Diffraction

G. Schönhense[1], K. Medjanik[1], S. Babenkov[1], D. Vasilyev[1], M. Ellguth[1,*], O. Fedchenko[1],

S. Chernov[1], B. Schönhense[2], and H.-J. Elmers[1]

[1] Johannes Gutenberg-Universität Mainz, Institut für Physik, Staudinger Weg 7, 55128 Mainz, Germany

[2] Imperial College, Department of Bioengineering, South Kensington Campus, London SW7 2AZ, UK

* now Surface Concept GmbH, Am Sägewerk 23a, 55124 Mainz, Germany

**Abstract**

Owing to strongly enhanced bulk sensitivity, angle- or momentum-resolved photoemission using X-rays is an emergent powerful tool for electronic structure mapping. A novel full-field $k$-imaging method with time-of-flight energy detection allowed rapid recording of 4D ($E_B, \mathbf{k}$) data arrays ($E_B$ binding energy; $\mathbf{k}$ final-state electron momentum) in the photon-energy range of 400-1700eV. Arrays for the $d$-band complex of several transition metals (Mo, W, Re, Ir) reveal numerous spots of strong local intensity enhancement up to a factor of 5. The enhancement is confined to small ($E_B, \mathbf{k}$)-regions ($\Delta k$ down to 0.01 Å$^{-1}$; $\Delta E_B$ down to 200 meV) and is a fingerprint of valence-band photoelectron diffraction. Regions of constructive interference in the ($E_B, \mathbf{k}$)-scheme can be predicted in a manner resembling the Ewald construction. A key factor is the transfer of photon momentum to the electron, which breaks the symmetry and causes a rigid shift of the final-state energy isosphere. Working rigorously in $k$-space, our model does not need to assume a localization in real space, but works for itinerant band states without any assumptions or restrictions. The role of momentum conservation in *Fermi's Golden Rule* at X-ray energies is revealed in a graphical, intuitive way. The results are relevant for the emerging field of time-resolved photoelectron diffraction and can be combined with standing-wave excitation to gain element sensitivity.

# 1. Introduction

Owing to the increased probing depth, angular- or momentum-resolved photoelectron spectroscopy in the X-ray range is rapidly gaining importance for electronic structure analysis of solids. The increased information depth facilitates access to the 3D electronic structure. True bulk sensitivity in the valence range has been proven using conventional spectroscopy [1-7] and *k*-microscopy [8] and photoemission in this regime has much potential. High-brilliance, high-resolution X-ray beamlines at Synchrotron sources and upcoming free-electron-laser sources, and advanced electron energy analysers with high performance in the hard X-ray range provide an excellent basis for future experiments. First photoemission experiments on samples with protective cap layers, buried layers in thin-film devices, in-operando devices, or samples reacting with a gas atmosphere have recently broken old paradigms of photoemission.

In the X-ray range, the wavelength of the excited photoelectrons is of the order of the atomic distances in the solid. Hence, photoelectron diffraction (PED - also referred to as XPD in the X-ray range) [9-15] influences the observed photoemission signals. This phenomenon is well understood for core-level photoemission, but thus far, data for PED in valence-band photoemission are sparse and results were interpreted analogously to core-level PED after integrating over a larger energy range [16-18]. It was found that the initial-state orbital angular momenta influence the valence-band PED signal [13].

PED/XPD in photoemission from core levels is a powerful method for gaining information about the geometrical structure of the photo-emitting atomic layers, surface reconstruction and relaxation; adsorbate sites; and distances (excellent overviews are given in [1,9-11,14]). By exploiting exchange scattering and multiplet splittings, even antiferromagnetic short-range order has been probed by PED [19,20]. Experimentally, PED is studied using angular-resolved photoelectron spectroscopy, usually by rotating the sample about its surface normal, but some studies have used a display-type electron analyser [21,22]. Core-level XPD is the result of a localized excitation at a given atomic site and the scattering of the resulting photoelectrons off neighbouring atoms. In early observations on single crystal surfaces the angular distributions were interpreted as being caused by reflection of the photoelectrons on lattice planes of the three-dimensionally periodic bulk crystal [23,24]. A two-beam dynamical theory was applied to explain the azimuthal variations of photoelectron intensities for single crystal copper [25,26]. However, at lower typical XPS energies of < 1.5 keV, short-range order scattering in a cluster has become the dominant mode of analysing PED data, e.g. [27]. The intensity variations, dominated by Kikuchi bands and single-atom forward scattering can be well reproduced using these models, but the two models being consistent with one another if fully converged [15]. As energy increases, the scattering becomes more forward peaked as a result of forward scattering at rows of atoms seen at typical energies of 1 keV. Cluster approaches [26] have proven to reproduce the experimental diffraction patterns with fair agreement for cluster sizes as small as few nm. A quantitative comparison [15] between the XPD cluster picture and dynamical electron scattering from lattice planes showed that the latter is more appropriate for very high energies.

Early studies for valence-band XPS in the high-energy, high-temperature, low $k$-resolution limit revealed matrix-element weighted densities of states (MEWDOS), modulated by XPD effects [17,18]. The first real hard X-ray ARPES experiment was performed by Gray et al. [1]. A two-step normalization process was used to eliminate the MEWDOS and XPD effects, in order to uncover the correct band dispersions. That work showed clear dispersions for W and GaAs samples at photon energies of 6 keV and 3.2 keV, respectively, in good agreement with one-step photoemission theory [28].

Angular- or momentum-resolved photoemission experiments in the X-ray range are hampered by strongly dropping photoemission cross sections and an increase of electron-phonon scattering with increasing photon energy. Photoelectron momentum microscopy constitutes a novel experimental ansatz to study valence-band photoemission at X-ray energies with enhanced detection efficiency. This method exploits the equivalence of the "spatial-frequency pattern" in the Fourier plane of an electron lens (e.g. the cathode lens of a photoelectron microscope) and the lateral $k$-distribution of the electrons emitted from a planar, solid sample. Winkelmann et al. were the first to apply this method for the study of photoelectron diffraction effects in a momentum microscope with a dispersive energy analyser. The authors observed the Mahan cone [29] and surface-barrier scattering [30] in Cu crystals at low energies (21.2 eV). In the present experiment, time-of-flight energy detection is used to record the $k$-distribution of the full $d$-band complex in a single measurement. The third momentum component (perpendicular to the surface) is accessed by varying the photon energy in the soft X-ray range. The ($E_B$,$k$) parameter space in the $k$-region of interest is mapped with a maximum degree of parallelization (for details, see [8]).

The present work was motivated by the appearance of strong intensity modulations caused by XPD/PED effects in valence-band mapping of various transition metals. Fig. 1 shows selected momentum images (isosurfaces at certain values of $E_{final}$) taken for the $d$-bands of Mo(110), W(110), Re(0001) and Ir(111). The data have been recorded using the ToF $k$-microscope described in [8] in the geometry sketched on top of Fig. 1 with circularly-polarized soft X-rays from beamline P04 of PETRA III (DESY, Hamburg). All images show the sum of two $k$-patterns taken for right- and left-circular polarization, thus eliminating the circular dichroism in the photoelectron angular distribution. The impact angle was 22° with respect to the surface plane; the sum of the two helicities corresponds to unpolarized light in near grazing incidence. The strong local character of the intensity modulations rules out that the enhancement regions are due to a photon-polarization effect.

In order to increase the Debye-Waller factor the samples were cooled to 40 K. All examples exhibit *pronounced local intensity enhancements* in small regions of the ($E_B$,$k$) parameter space ($\Delta k$ few hundredths of Å$^{-1}$, $\Delta E_B$ few hundred meV), overlaid on the valence-band patterns. The intensity distributions show neither Kikuchi bands nor the signature of forward scattering from atom rows nor do they reflect the crystal symmetry. We will show below that the local enhancements (marked by dashed ellipses) result from PED and that the lack of symmetry is a fingerprint of the transfer of photon momentum to the photoelectron. Fig. 1 is apparently in conflict with literature XPD data showing perfect crystal symmetry. The reason for this apparent contradiction lies in the different data acquisition mode: In the present experiment sample, detector and photon beam are fixed, whereas in conventional experiments the sample is rotated about its surface normal. In Section 3, we will discuss the findings in detail. Here, we mention just the most surprising facts, i.e. the strong local confinement, the missing symmetry and the dramatic changes of the intensity modulations with energy: All patterns in the first row (Fig. 1(a-e)) have been taken for Mo at fixed photon energy (hν= 1700 eV) but at different final-state energies as stated in the panels (energy resolution ~80 meV). The small increments of 1.5 eV (corresponding to steps of only $\Delta k \sim 0.01$ Å$^{-1}$ in the final-state momentum vector) lead to a puzzling multitude of irregular local enhancements, essentially without any visible systematics.

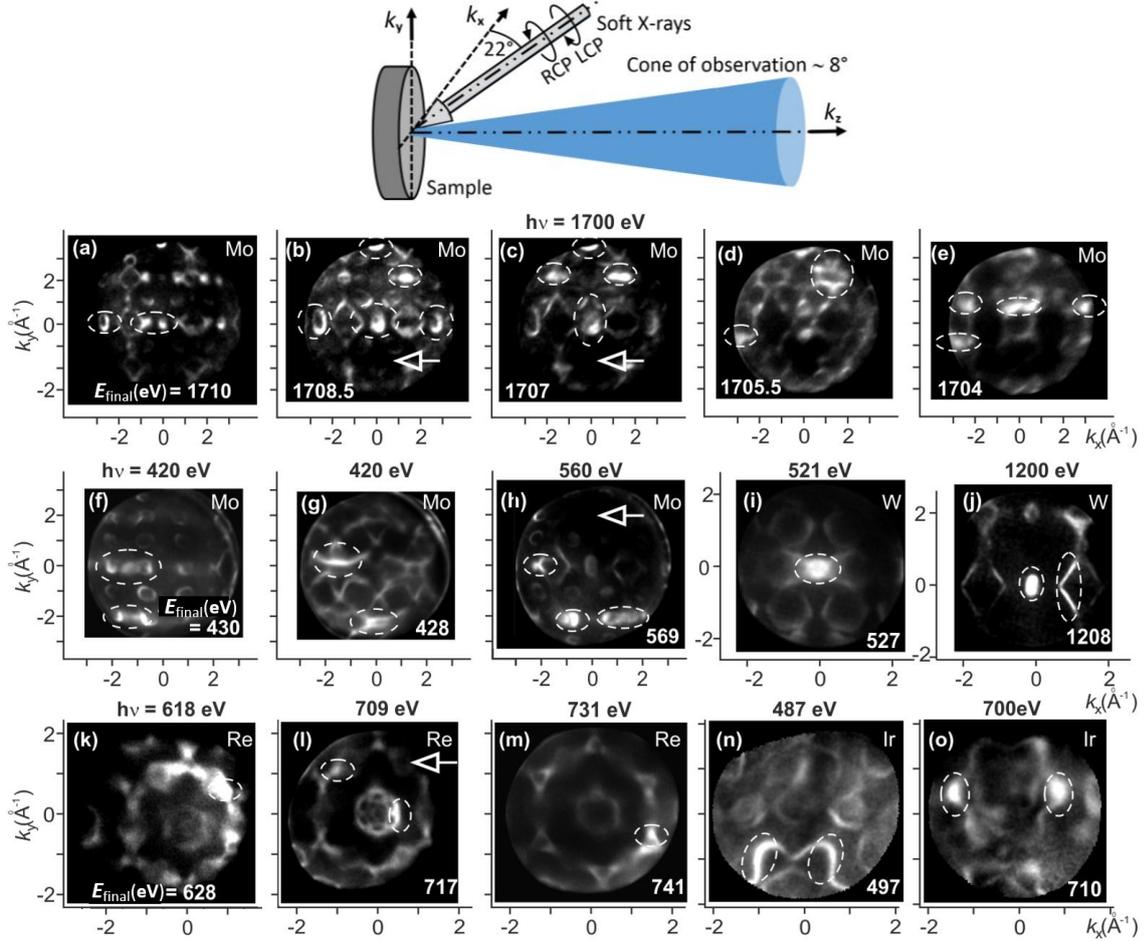

**Figure 1.** Appearance of valence-band photoelectron diffraction in *k*-microscopy as an "irregular" pattern of local intensity modulations. Top panel, geometry of the experiment. The momentum distributions (energy isosurfaces) were taken for Mo(110) (a-h), W(110) (i,j), Re(0001) (k,l,m) and Ir(111) (n,o) at various photon energies and final-state energies as denoted in the panels. In all panels, areas of local intensity enhancement (dashed ellipses) and extinctions (arrows) appear, confined to certain *k*-regions and energies. Note the strong variations in (a-e), all taken at hν= 1700 eV but at 5 different final energies $E_{final}$, separated by increments of only 1.5 eV. $E_{final}$ refers to the photoelectron (kinetic) energy inside the solid, before passing the surface barrier.

The goal of the present work was to find a suitable description of XPD/PED in *k*-microscopy of valence bands in a region of parameter space that is not dominated by Kikuchi bands, yet is also beyond conventional cluster calculations, which are not easy to interpret. We adopt the acronym VBPED for *valence-band* PED from [13], emphasizing that its origin differs from the conventional PED, which deals with a fully localized core-level emitter, and can be properly described in a real-space cluster model, even though this core-like picture has been used to describe MEWDOS valence PED previously [17,18]. The diameter of the *k*-region in Figs. 1(a-

h) is 6 Å$^{-1}$, corresponding to an observed cone with half angle of only 8° for 1700 eV (first row). At such small polar angles, forward scattering from atom rows along off-normal high-symmetry directions cannot explain the results (cf. Fig. 9 of [10]). Below, we propose a model for a quantitative analysis of the VBPED patterns based on Umklapp processes involving reciprocal lattice vectors **G**. A graphical representation resembling the Ewald construction in conventional diffraction allows for a quantitative prediction of regions of constructive interference in the ($E_B$,**k**)-scheme. An important fact is the breaking of symmetry caused by the photon momentum $k_{h\nu}$ being fully transferred to the photoelectron, as discussed in previous soft- and hard- X-ray ARPES [1-4]. In the photon-energy range from 400 to 1700 eV, $k_{h\nu}$ increases from 0.20 to 0.86 Å$^{-1}$ and this causes a substantial shift of the entire momentum pattern. The effect is visible in the downward shift of the momentum scale of Fig. 1 panel (j) in comparison with panel (i), taken at identical adjustment but for h$\nu$= 1200 and 521 eV, respectively (photon impact from top to bottom).

The results are of general importance for *k*-microscopy experiments in the X-ray range, because the momentum distributions observed at a given photon energy are strongly modulated by VBPED. Fig. 1 also reveals regions where the band features appear attenuated (marked by arrows), this might be a hint on conditions of destructive interference. If disregarded, this substantial influence of VBPED on the observed band features can cause a misinterpretation of the observed intensity with respect to the spectral density of states.

## 2. Valence-band photoelectron diffraction described in *k*-space
### *2.1 Direct transitions in periodic k-space*

Within the framework of first order time-dependent perturbation theory, the photoemission intensity can be derived from Fermi's golden rule describing the transition probability *W* from an initial state $\varphi_i$ to a final state $\varphi_f$,

$$W = \frac{2\pi}{\hbar} | <\varphi_f| \Delta | \varphi_i > |^2 \, \delta(E_f - E_i - h\nu) \qquad (1)$$

with the perturbation operator $\Delta$ representing the electromagnetic field of the light including its polarization state. The $\delta$–function accounts for energy conservation. Momentum conservation in case of non-negligible photon momentum is discussed below using a graphical intuitive model instead of a second $\delta$-function in Eq. (1).

Considering the correct final state, the one-step model describes the actual excitation process, the transport of the photoelectron to the crystal surface as well as the escape into the vacuum as a single quantum-mechanically coherent process including all multiple-scattering events. Although numerical implementations of this model predict experimental data with increasing accuracy [31], the aspect of photoelectron diffraction still requires a large numerical effort. Furthermore, the numerical simulation does not allow a direct insight into the diffraction paths. Illustrative approximations for the case of core-level photoelectron diffraction have been successfully developed in a real space representation, considering localized spherical initial states [9]. A similarly descriptive representation for the case of valence-band photoelectron diffraction demands a *k*-space model (if no *ad-hoc* assumptions about an initial-state localization are made).

In the *E*-vs-***k*** scheme, photoemission in the soft and tender X-ray regime is described by direct transitions into *quasi-free-electron-like* final states, see e.g. [32,33]. This means that the electrons show a parabolic dispersion of the final-state energy $E_{final}$ vs final-state momentum $k_f$, but their effective mass $m_{eff}$ can still differ somewhat from the free electron mass $m_e$. For tungsten we found $m_{eff}/m_e$ = 1.07 at 1000 eV [8] whereas this value reduces to 1 at 6 keV as observed by Gray et al. [1]. For molybdenum we found $m_{eff}/m_e$ = 1 already at 1700 eV [37]. At low energies the final-state band deviates from parabolic dispersion. For rhenium at a photon energy of 15 eV we find $m_{eff}/m_e$ = 1.22 as will be shown in Section 3.3.

Besides the effective mass, the absolute energy position of the final-state parabola is an empirical quantity. We write the dispersion relation as

$$k_f = (1/\hbar)\sqrt{2m_{eff}E_{final}} \quad \text{with} \quad E_{final} = h\nu - E_B + V_0^* \qquad (2).$$

This equation looks different from the conventional description [32,33], because we refer the inner potential $V_0^*$ to the Fermi energy and not to the vacuum level. In this approximation, assuming a transition in periodic *k*-space, there is no surface, no workfunction and no refraction of the outgoing electron wave at the surface barrier (which would involve the work function and $V_0$). Diffraction effects (Umklapp) happening at the surface are excluded.

The assumption of a parabolic final-state dispersion as parametrized in Eq. (2) deserves some additional considerations in the context of *k*-microscopy. The momentum microscope is like a magnifying glass looking directly into *k*-space on a linear achromatic $k_\parallel$ scale. As we will see below, we can localize high-symmetry points in *k*-space very precisely. The full vector ***k***$_f$ can be quantitatively determined with the precision of the lattice constant of a material (because the reciprocal lattice is known with this precision). The only precondition at this stage is that the reciprocal lattice is periodic. Several cases of absolute ***k***-determination will be shown below. A particularly simple example is the low-energy case in Section 3.3 since the photon momentum is negligible. The bright spot in Fig. 7(h) reveals that the final-state momentum vector is exactly ***k***$_f$ =(0,0,2.82) Å$^{-1}$. Upon increasing $h\nu$ the momentum $k_f$ increases and the slope gives the effective mass according to Eq. (2). However, the bottom of the final-state parabola is not known. Moreover, $m_{eff}$ and in the general case also $V_0^*$ must be assumed to be energy dependent. Hence, there is no "universal" final-state parabola that is valid throughout a large energy range. Rather, we consider $V_0^*$ as an empirical fit parameter as well.

As further (more technical) motivation for referring the inner potential to the Fermi energy we recall a special property of *k*-microscopes: the refraction step at the surface drops out since the instrument records momentum distribution patterns directly in the $k_\parallel$ coordinate system *inside* of the material. This $k_\parallel$ scale is constant independent of the kinetic energy and the work function [34]. Work function changes only lead to a change of the diameter of the photoemission horizon, but not of the $k_\parallel$ scale (except for a small effect of the chromatic aberration of the lens system). For the elements investigated here, we assume an inner potential of $V_0^*$= 10 eV (corresponding to about 15 eV when referred to the vacuum level).

Since we can measure the final-state momentum in a parameter-free manner and since diffraction is readily described by a transfer of momentum vectors, it is close to being able to put valence-band PED into a *k*-space model as well. In the periodic zone scheme, each Brillouin zone contains the full set of valence bands. The dependence on binding energy $E_B$ adds the fourth coordinate. The full 4D spectral density consists of occupied and unoccupied regions, separated by the Fermi surface. Analogous to the discussion of the Fermi surface, we have a multitude of surfaces in 3D $k$-space which separate occupied from unoccupied states/regions for a given $E_B$. It is descriptive to visualise the band structure as a set of such bounding surfaces. Stated differently, the full 4D density has a 3D bounding volume of occupied regions in this 4D space. Different energies then give different 2D cuts (which we term "energy isosurfaces") through this volume, one of which is again the Fermi surface. The $E_B$=const. surfaces are often fragmented into isolated electron and hole pockets and appear as periodically-repeated patterns, identical in all BZs. This notion is different from the conventional description in terms of $E$-vs-$k$ plots for certain high-symmetry $k$-directions. As we will see below, it is a very convenient basis of understanding diffraction of photoelectrons originating from a propagating Bloch wave because the initial $k$-vector is accounted for in the momentum balance.

In the photoexcitation process, photon energy $h\nu$ and photon momentum $k_{h\nu}$ (being significant in the X-ray range) are both transferred to the photoelectron. Energy conservation demands that all final states of the photo-transition are located on a sphere with radius $k_f$ given by Eq. (2). Momentum conservation causes a shift of the centre of the final-state sphere from the origin $k$=(0,0,0) by the vector $k_{h\nu}$. Figure 2 shows a quantitative scheme of a photoexcitation in molybdenum at a photon energy of 400 eV. The transition leads to the 4$^{th}$ repeated BZ. Symmetrized periodic background patterns like the one in Fig. 2(a,b) are extracted from measured 4D arrays: We first select the proper isosurface from the array and then cut this isosurface in the relevant plane (in this case $k_z$-$k_x$). Since one of the axes is the perpendicular momentum $k_z$, these patterns are not just measured $k$-images but they result from measuring $k_x$-$k_y$ patterns at many photon energies, concatenating them along $k_z$ and then make a $k_y$=0 cut from the concatenated 4D array. Fig. 2 shows such cuts through two measured isosurfaces (from [37]), shown in the insets: the Fermi surface (a) and the surface at $E_B$= 1 eV (b). The advantage of this representation is that it immediately shows all the initial $k$-vectors in this plane corresponding to a certain energy.

Since in general the photon momentum has an arbitrary direction with respect to the reciprocal lattice of the sample, plots like in Fig. 2(a,b) must consider full 3D $k$-space. Here we have chosen a photon impact direction in the $k_x$-$k_z$ plane, like in the experiments shown below. The 4D character of the photoemission process is evident: the photon-energy dependence for a given binding energy $E_B$ (Fig. 2(a)) and the binding-energy dependence for a given photon energy (Fig. 2(b)) both lead to a set of different final-state spheres. In addition, the shape of the isosurface in the periodic band-structure pattern changes rapidly with binding energy, cf. examples for $E_B$= 0 and 1 eV in the insets of Figs. 2(a) and (b).

Here, we use the approximation that the effective mass does not depend on $k$. In general, the sphere might be slightly deformed if the effective mass is direction-dependent. Since we will refer to it in the next section, Fig. 2(c) shows the Ewald sphere, a visual interpretation of the Laue equations in a diffraction experiment.

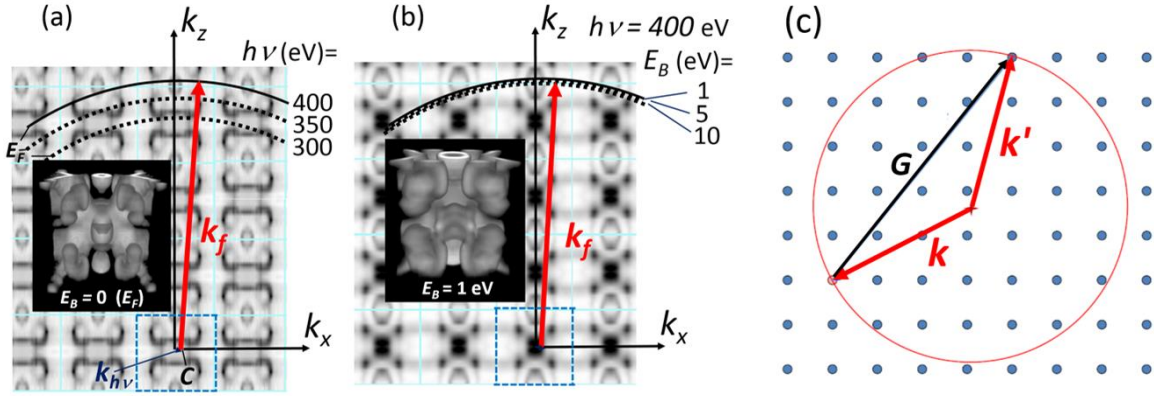

**Figure 2.** Model of photoemission into quasi-free-electron-like final states. Owing to energy conservation, the final states are located on a sphere of radius $k_f$. This radius depends on photon energy hν (for fixed $E_B$) (a) and on binding energy $E_B$ (for fixed hν) (b). The centre C of the sphere is displaced from the origin $\mathbf{k}=(0,0,0)$ by the vector of the photon momentum $\mathbf{k}_{h\nu}$. Plots (a) and (b) are to scale for Mo(110) at a photon energy of 400 eV. The background patterns are cuts (at $k_y$=0) of the periodically-repeated measured 4D spectral-density array for the Fermi energy $E_F$ (a) and for a binding energy of 1 eV below $E_F$ (b); dark denotes high spectral density. Dashed squares denote the 1st BZ. The insets are 3D views of the measured isosurfaces; note the strong change in shape despite the small energy difference between (a) and (b). For comparison, (c) shows the Ewald-sphere construction in electron diffraction.

The basis of this description of quasi-free-electron final states in photoemission is the following analogy:

*A <u>diffraction spot</u> occurs whenever the Ewald sphere intersects a reciprocal lattice point; a <u>photoemission signal</u> occurs wherever the final-state sphere (displaced from the origin by $\mathbf{k}_{h\nu}$) intersects a band feature in one of the repeated Brillouin zones.*

The Ewald sphere does not give information on the intensity of diffraction spots or on systematic extinctions. Likewise, the intersection regions of the final-state sphere with the periodic pattern of isosurfaces only show where band features are visible in principle. The actual intensity of an observed band depends on the matrix element in Eq. (1) that also accounts for the photon polarization. The correct transition matrix element (without restrictions or approximations) thus includes the selections rules due to the photon polarization and multiple scattering in the final-state wavefunction $\varphi_f$ (the "time-reversed LEED state" [31]). Hence, the matrix element contains the information on all possible diffraction paths in the total final state of the outgoing photoelectron. In order to understand the origin of a particular intensity enhancement we will employ a kinematic model, which predicts the possible VBPED spots in a given reciprocal lattice.

### 2.2 The 1D case of normal emission

In order to understand how XPD/PED is observed in a *k*-space microscope we express the diffraction conditions in 4D ($E_B$,$\mathbf{k}$) parameter space. In bulk-sensitive valence-band photoemission, all BZs contain the identical set of energy surfaces as shown in Figs. 2 (a,b). We begin with the simplest case, i.e. normal emission, corresponding to Bragg scattering at lattice planes parallel to the surface. For constructive interference the phase difference of the outgoing scattered and direct partial waves must be an integer multiple of $2\pi$. For forward scattering the spacing $d_z$ of the atom planes parallel to the surface must coincide with an integer multiple of the wavelength $\lambda$ of the photoelectrons. In backward scattering the path difference is twice the spacing $d_z$. Full (half) integers originate from constructive interference in forward (backward) scattering in *normal emission*:

$\qquad d_z = n\,\lambda \qquad$ *forward scattering* $\qquad$ (3),

$\qquad d_z = \dfrac{n}{2}\lambda \qquad$ *backward scattering* $\qquad$ (4).

With the reciprocal lattice vector perpendicular to the surface $G_z = 2\pi/d_z$ and $k_f = 2\pi/\lambda$ these relations translate into *k*-space as

$\qquad k_f = n\,G_z \qquad$ *forward scattering* $\qquad$ (5),

$\qquad k_f = \dfrac{n}{2} G_z \qquad$ *backward scattering* $\qquad$ (6).

The second equation corresponds to the Bragg condition $2d \sin\theta = n\lambda$ for $\theta = 90°$ (the angle is defined with respect to the atomic plane). According to Eqs. (5,6) the condition for constructive interference in the direction perpendicular to the surface is that the *final-state sphere intersects the centre (or the boundary) of a Brillouin-zone* as sketched in Figs. 3(a) and (b), respectively. For the experiments shown below the angle of incidence is 22° from the surface. Hence, the main part of the shift of the sphere by the photon momentum acts in transversal direction. The *k*-microscope records the intensity pattern on a spherical section with diameter up to 8 Å$^{-1}$ close to normal emission, sketched as "red cap" in Figs. 3 (a,b) (for details, see [8,35]).

We have tested relation (5) for the Mo(110)-surface and indeed found a pronounced forward-scattering maximum at h$\nu$= 460 eV, where the final-state vector just reaches the centre of the 4$^{th}$ repeated BZ (see Fig. 4 and discussion in the next Section). Destructive interference occurs when the phase-shift difference is $\dfrac{2n-1}{2}\pi$ and can lead to an attenuation of band features. Regions of attenuation are visible in Fig. 1, denoted by arrows.

Relations (5,6) have important consequences in 3D electronic structure mapping using *k*-microscopy as performed in [8,35-37]. The regions in the centre and at the boundary of the BZ in the direction perpendicular to the surface appear with enhanced intensity. Regions where the condition for destructive interference is fulfilled appear with too low intensities, so that band features can be missing. These effects may also show up as discrepancy in comparison with photoemission calculations that do not include VBPED. Note that this model is conceptually different from the conventional description of core-level PED, which assumes a localized initial state. Our momentum-transfer model works rigorously in *k*-space and is applicable to itinerant band states without any assumptions or restrictions. The only necessary precondition is a periodic band pattern in momentum space (demanding lattice periodicity in real space).

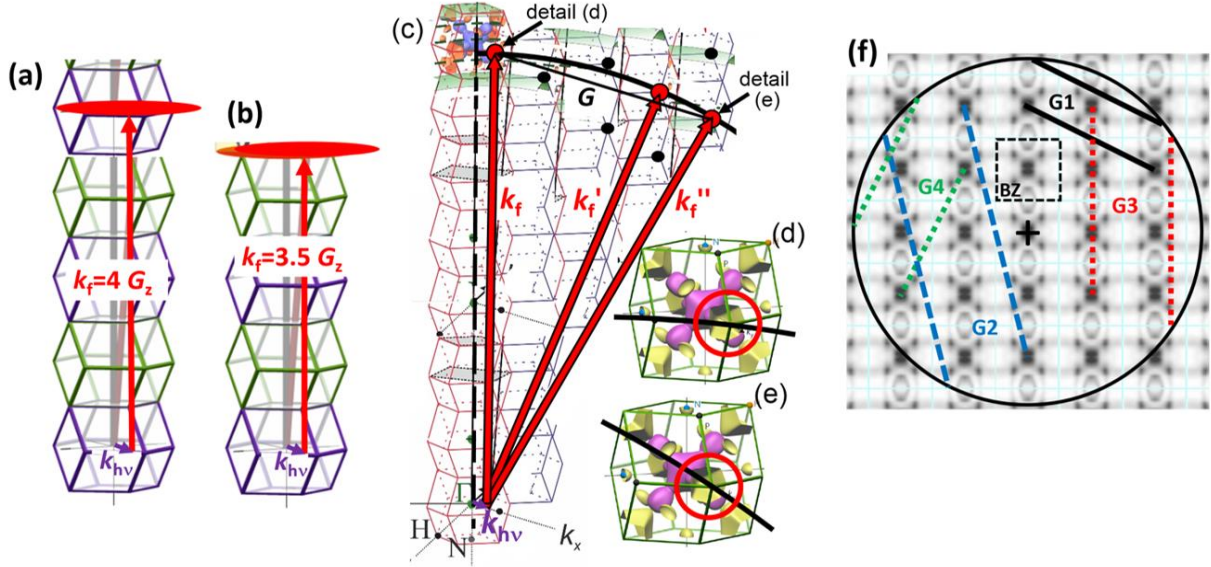

**Figure 3.** Direct transitions fulfilling conditions for constructive interference in normal emission due to forward (a) and backward scattering (b). The general case (c) resembles the Ewald construction, a graphical way to find "spots" where the Laue condition ($k_f' - k_f = G$) is fulfilled. Details (d) and (e) mark ($E_B, k$)-sectors, where the final-state sphere intersects identical regions in different repeated BZs. Vectors $k_f$, $k_f'$ and $k_f''$ are located on the same final-state sphere and reach equivalent points. Reciprocal lattice vectors $G$ give their distances. (f) Scheme illustrating that all reciprocal lattice vectors inside of the final-state sphere can be shifted so that both ends lie on the sphere. Experimental background pattern as in Fig. 2(b), photon momentum neglected; BZ marks the Brillouin zone.

### *2.3 The 3D case – valence-band PED in periodic momentum space*

Forward scattering is the most easily explained component of VBPED, but only captures a small fraction of all possible scattering processes involving arbitrary reciprocal lattice vectors $G$ of 3D momentum space. The task is to find intersection points of the final-state sphere (displaced by the photon momentum) with equivalent $k$-points in different repeated BZs. Such cases are illustrated in Fig. 3(c), where the vectors $k_f'$ and $k_f''$ reach the same band feature as vector $k_f$, but in different BZs. The red circles in insets (d) and (e) mark the ($E_B, k$)-regions, where the final-state sphere intersects such identical points. Their distances are given by certain reciprocal lattice vectors $G$. Energy conservation demands that both ends of $G$ must lie on the final-state sphere (cf. Fig. 3 (c,f)), which defines the ($E_B, k$)-region.

The generalization of eqs. (5,6) in 3D $k$-space is the *Laue equation*:

$$k_f = k_f' - G \qquad (7)$$

In fact, each reciprocal lattice vector inside the final-state sphere corresponds to a specific ($E_B, k$)-region which is intensified by VBPED, cf. Fig. 3(f), where we have chosen the $E_B = 1$ eV isosurface, like in Fig. 2(b). When a band feature crosses this region, its intensity is modulated. *The vector $G$ shifted to touch the sphere on both ends defines the sector for constructive interference, whereas the isosurface crossing these points defines at which binding energy the enhancement appears.* In this respect, PED in valence-band photoemission differs from conventional diffraction, where the Ewald sphere must intersect a reciprocal lattice point, as sketched in Fig. 2(c). In our model, the momentum of the Bloch wave of the initial state is included in the total momentum balance. In Fig. 3(f) it is defined by the equivalent points on the periodic energy isosurface in *k*-space, which are connected by the vectors **G** on the sphere. In this way, no *ad-hoc* assumption on a localization in real space is required. The number of ($E_B$,**k**)-regions in which the Laue condition is fulfilled increases with the number of BZs on the surface of the sphere which is a linear function of photon energy. The radius of the sphere increases with the root of the energy and the surface area is proportional to the square of the radius. In the sequence VUV, soft, tender and hard X-rays, for hν = 50, 400, 1700 and 6000eV, the numbers of BZs on the surface of the sphere are about 50, 190, 780 and 2800, respectively (taken the parameters from the example of Fig. 2). The normal-emission case in Eqs. (5,6) corresponds to |**G**| = 0 (forward scattering) and |**G**| = 2|$k_f$| (backward scattering).

Being based on the Laue condition, this model represents the kinematic approximation, i.e. multiple scattering is neglected. This approximation is good for energies in the X-ray range, whereas at low energies multiple scattering becomes significant, thus increasing the number of ($E_B$,**k**)-regions with enhanced intensities. Moreover, these considerations are valid for an infinitely large lattice. In practice, the inelastic mean free path (IMFP) [38] limits the path lengths in real space. In *k*-space this corresponds to a relaxation of the exact diffraction condition leading to a reduction of the intensity enhancement and broadening of the profiles of the diffraction features. In addition, the Debye-Waller factor leads to a temperature-dependent weakening of the diffraction features and increase of diffuse scattering [1,2]. This factor decreases exponentially with increasing temperature and with increasing modulus |**G**|. In the measurements shown below the sample temperature was 40 K, where the Debye-Waller factor is still rather large. Finally, the atomic form factor of the scattering atoms also modifies the amplitudes of the diffracted partial waves. The atomic form factor also induces an additional phase shift of the wavefunction, depending on the wavelength. However, in the energy range used in the present study this form factor contribution and atomic phase shift are negligibly small [10], but they become significant for higher kinetic energies.

### 3. Experimental results and quantitative analysis

*3.1 Near-normal emission*

The present results have been taken using the ToF *k*-microscope described in [8], using the geometry sketched on top of Fig. 1. In order to validate Eq. (5), we have looked for intensity enhancement in the centre of the *k*-distributions. A prominent case of forward scattering in normal emission occurs for the Mo(110)-surface at a photon energy of 460 eV, see Fig. 4. We observe strong intensity enhancement in the centre of the momentum patterns at $E_F$ (a) and 2 eV below $E_F$ (b), also visible in the $E_B$–$k_y$ section (c). In comparison, sections away from normal emission show more homogeneous intensity distributions (d,e). The strongest intensity enhancement occurs in the centre (i.e. for normal emission) at a binding energy of $E_B \sim 1.7$ eV ($E_{final}$= 468.3 eV) as visible in the intensity profile Fig. 4(f) taken in the marked rectangular area in the centre of (b). Fig. 4(g) shows a quantitative plot of the transition at hν= 460 eV in the $k_z$-$k_x$ plane. The origin of the periodic pattern is discussed in Fig. 2. The curvature of the final-state sphere with radius $k_f \sim 11.6$ Å$^{-1}$ on the scale of one BZ (radius ~1.4 Å$^{-1}$) is rather small. The shift due to the photon momentum is small as well; the resulting tilt angle of $k_f$ with respect to the $k_z$-axis is only 1.1°.

The perpendicular reciprocal lattice vector of Mo(110) has a length of $G_z=G_{hkl}=G_{110}=2.824$ Å$^{-1}$. The inner potential (referred to $E_F$) is $V_0^* \sim 10$ eV and the effective mass is $m_{eff} = 1.05\,m_e$ as derived from 4D band mapping [37]. At 460 eV, the photon momentum is 0.23 Å$^{-1}$ and its perpendicular and in-plane components at the shallow photon impact angle of 22° from the surface are $k_{h\nu}^z = 0.09$ Å$^{-1}$ and $k_{h\nu}^x = 0.21$ Å$^{-1}$. For the kinetic energy of 468.3 eV Eq. (1) yields $k_f$ = 11.34 Å$^{-1}$; for normal emission $k_f^z = k_f - k_{h\nu}^z = 11.25$ Å$^{-1}$ = 3.98 $G_{110}$. Hence, this perpendicular wave vector leads to the centre of the 4$^{th}$ repeated BZ along $k_z$ and proves Eq. (5) with **$k_f$** = 4 **$G_{110}$**. For this particular case of normal emission, a calculation for the Mo(110) surface in the localized model using the EDAC code [27] also shows an intensity maximum close to 460 eV and furthermore reveals additional oscillations of intensity with energy due to higher-order interferences and multiple scattering.

Such measurements provide a metric in *k*-space for the determination of the centres and boundaries of repeated BZs. It should be mentioned, however, that the parameters $V_0^*$ and $m_{eff}$ in Eq. (2) depend on energy (for details, see [8]). For molybdenum at h$\nu$= 1700 eV, we find $m_{eff} \approx m_e$, in agreement with the result for tungsten at 6 keV [1].

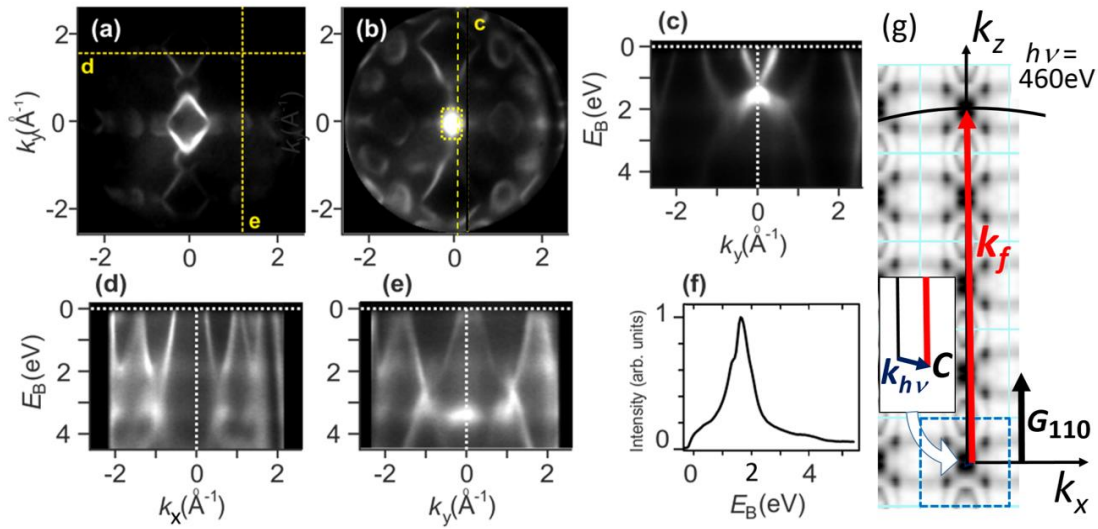

**Figure 4.** (a-c) Pronounced forward scattering in normal emission for valence photoelectrons from the Mo(110) surface at h$\nu$= 460 eV, visible in the isosurfaces at $E_F$ (a), $E_B$= 2 eV (b) and in a $E_B$–$k_y$ section (c) along the dashed line in (b). Away from the interference condition, the intensity is essentially evenly distributed as seen in sections $E_B$–$k_x$ (d) and $E_B$–$k_y$ (e), cut along the dashed lines in (a). The intensity enhancement is quantified by the intensity-vs-$E_B$ profile (f), taken from the dotted rectangle in (b). (g) Scaled plot of the transition for Mo(110) at 460 eV in the $k_z$-$k_x$ plane. The final-state sphere intersects the centre of the 4$^{th}$ repeated BZ (diffraction condition $k_f$ = 4 **$G_z$**). Diameter of the *k*-field of view in (a) and (b) ∼ 5 Å$^{-1}$.

Away from the interference condition, the intensity is much lower and evenly distributed along the bands. As visible in sections (d) and (e), there are intensity enhancements across the entire field of view at certain energies (here at ∼1.8 and 3.3 eV). These horizontal "stripes" of enhanced intensity depend strongly on sample temperature and are a fingerprint of *quasi-elastic scattering at phonons*. Their energy position corresponds to maxima in the matrix-element weighted density of states (MEWDOS) as was observed and discussed in earlier work [1,16-18]. The cross section for electron-phonon scattering increases with kinetic energy and becomes the dominant "loss channel" in band mapping at high energies. Phonons can transfer large momenta in the scattering process, leading to randomization of the angular pattern. However, the corresponding energy transfer is limited to the 100 meV range (as visible in panels (d,e)), hence the use of the term *quasi-elastic*. Similar to diffraction experiments, the relative weight of the phonon-scattering channel depends on the Debye-Waller factor.

In 4D *k*-space mapping via scanning of the photon energy as described in [8,35-37], enhancement of the photoelectron intensity along the $k_z$-axis was indeed observed, whenever the final-state sphere crossed a BZ centre or boundary. An example for backward scattering is shown in Fig. 1(j). Here the hole pocket at the N-point of tungsten appears strongly enhanced by VBPED at hν= 1200 eV, see the bright oval shifted downward from the image centre by the photon momentum. Unlike Fig. 4 this corresponds to the boundary of the BZ. A photoemission calculation predicts a much lower intensity of this feature at this photon energy [36].

The example in Fig. 4 underlines an important property of valence-band PED: the *continuous* energy distribution (we recorded an energy band of ~6 eV) increases the probability to observe diffraction features in the restricted *k*-region viewed by the microscope. At hν= 460 eV the maximum of constructive interference occurs not at the Fermi energy but 1.7 eV below $E_F$. With increasing binding energy the kinetic energy and hence $k_f$ is reduced, according to Eq. (2). This means, if the wave vector of the electrons from the Fermi level is too large for constructive interference, there are electrons at a certain value of $E_B$, which just fulfil the diffraction conditions Eqs. (5-7). In this respect, photoelectron diffraction of valence electrons and (monoenergetic) core electrons fundamentally differ from each other. The situation in the valence range resembles X-ray diffraction with a white beam.

### *3.2 Diffraction involving arbitrary reciprocal lattice vectors*

For all materials studied in the soft and tender X-ray range we observed the, at first sight puzzling, multitude of irregularly distributed regions of intensity enhancement, as in the examples in Fig. 1. Symmetric VBPED patterns like Figs. 1(i,j) and 4(a,b) are rather an exception. A systematic analysis revealed that the general behaviour can be explained using the concept developed in Section 2. Figure 5 shows the analysis for the Mo(110)-surface at a photon energy of 1700 eV. At this high energy $k_f$ is 21.17 Å$^{-1}$, corresponding to the 8$^{th}$ repeated BZ. The photon momentum $k_{hν}$ = 0.86 Å$^{-1}$ causes a significant shift of the final-state sphere by 45% of the BZ radius. The diameter of the *k*-field of view in Fig. 5(a) is ~ 6 Å$^{-1}$, hence the first and four next BZs are visible. In real-space coordinates, this *k*-field corresponds to a cone with polar angle interval from 0°- 8° only. For PED this is a very small angular interval around normal emission where no Kikuchi lines and no direct paths for off-normal forward scattering are present. Momentum microscopes often focus on one BZ, whereas conventional PED experiments record typically 0°- 60°.

The $k_x$-$k_y$ cut (a) and $E_B$-$k_x$ cut (b) show pronounced intensity enhancement in a small region within 200 meV from the Fermi energy, whereas no enhancement is visible in the $E_B$-$k_y$ cut (c). Panel (d) shows a quantitative analysis: The final-state sphere runs through the 8$^{th}$ repeated BZ along $k_z$. It crosses the identical feature for $k_f' - k_f = G_{-1\,1\,0}$ (detail (e)) and $k_f'' - k_f = G_{-2\,14\,0}$.

Vectors G are labelled by the Miller indices. The photon momentum $k_{h\nu}$ causes a significant shift of the sphere (detail (f)). At this high energy we find $m_{eff} \approx m_e$.

In this case *both constructive interference conditions*, namely **G**$_{-1\,1\,0}$ and **G**$_{-2\,14\,0}$, lead to the observed strong enhancement feature indicated in Fig. 5(a,b). In the language of a diffraction experiment, the low- and high-index cases **G**$_{-1\,1\,0}$ and **G**$_{-2\,14\,0}$ belong to the zero-order and a higher-order Laue-zone, respectively. The high Miller index of 14 results from our choice of the $k_x$- and $k_y$-axes along [110] and [1-10], as appropriate for the (110)-surface and our observation geometry. Note that here we consider only one quadrant in a planar cut in the $k_x$-$k_y$ plane. There are more such conditions when the full Ewald-like sphere in 3D **k**-space is considered.

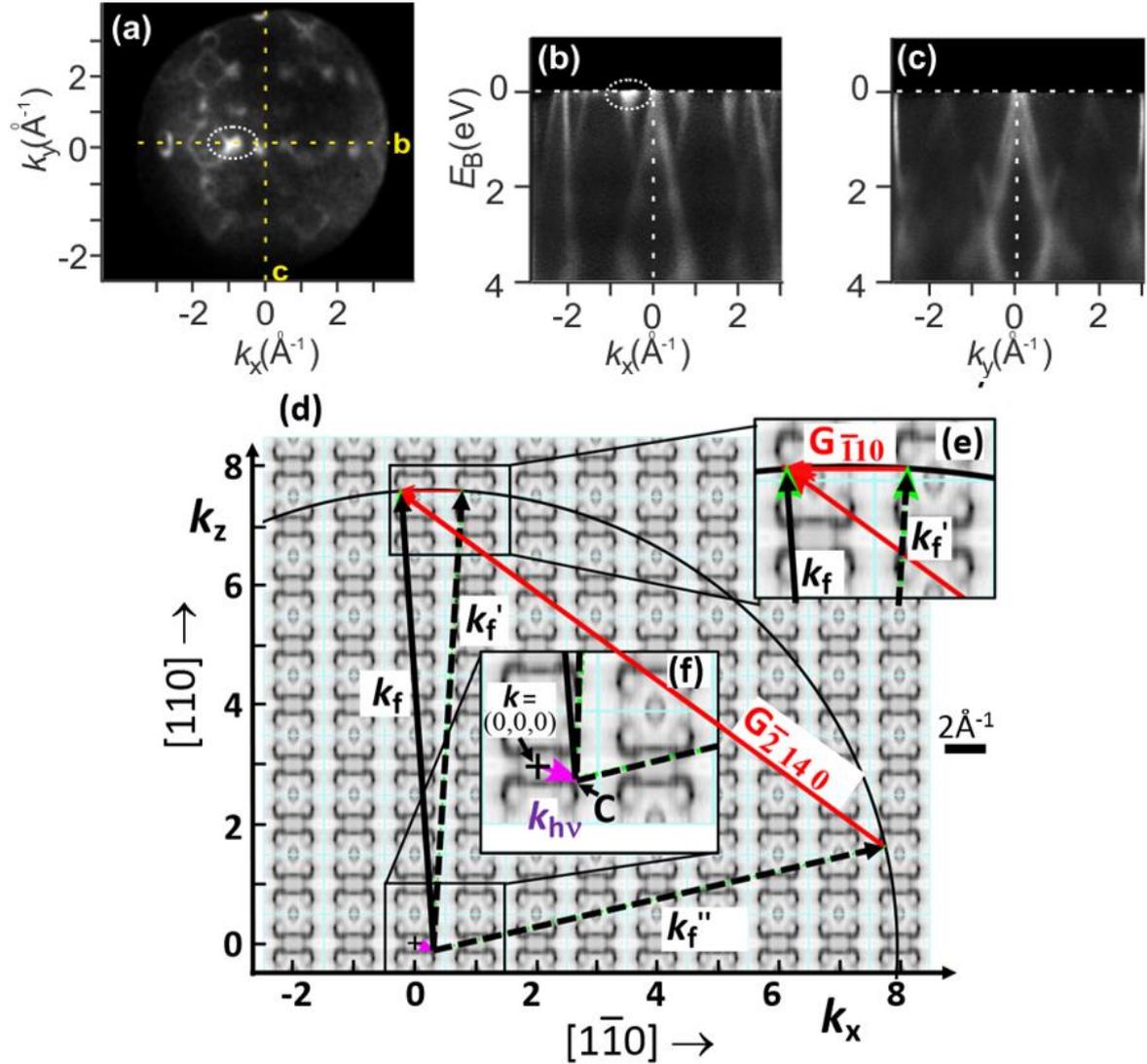

**Figure 5.** Valence-band photoelectron diffraction for the Mo(110) surface at h$\nu$= 1700 eV. An intensity enhancement is visible in a small local region at the Fermi energy in the $k_x$-$k_y$ cut (a) and $E_B$-$k_x$ cut (b), but it is absent in the $E_B$-$k_y$ cut (c). (d) Quantitative analysis: the background pattern is the periodically-repeated measured 4D array, cut at $E_F$ and $k_y = 0$ (dark is high spectral density). The final-state sphere runs through the 8$^{th}$ repeated BZ along $k_z$ and crosses the identical feature for $k_f' - k_f = $ **G**$_{-1\,1\,0}$ (cf. detail (e)) and $k_f'' - k_f = $ **G**$_{-2\,14\,0}$ (G's are labelled by the Miller indices). The photon momentum $k_{h\nu}$ causes a strong shift of the centre C of the sphere (cf. detail (f)).

There are several possible different views, which each reveal different aspects of the electronic structure. Compared to conventional $E$-vs-$k$ representations, Figs. 2 and 5 no longer clearly show band dispersion, but instead reveal how a total final photoemission state of a given energy is composed in k-space. Moreover, from such figures it is evident that a photoelectron from an itinerant initial state (Bloch wave with wave vector $k$) can undergo diffraction by transfer of a vector $G$ in the final state.

The results of Fig. 5 have been recorded at the onset of the so-called tender X-ray range, where prominent PED/XPD effects are to be expected. We found equally strong VBPED features also at lower energies in the soft X-ray range as demonstrated for the special case of normal emission in Fig. 4. An example taken at hv= 400 eV for VBPED with arbitrary vectors $G$ is shown in Fig. 6. The momentum distribution at $E_F$ (a) and the $E_B$-$k_y$ sections (e-l) show a number of regions with enhanced intensities. Panel (a) is dominated by cuts through electron and hole pockets, which appear as oval features with either inward or outward dispersion, respectively. The dispersion behaviour is visible in sections (e-l). Regions of enhanced intensity are located mostly in the 2$^{nd}$ BZ at negative $k_x$ values, see intensity plots (b) and (c), and in the 2$^{nd}$ BZ at negative $k_y$, see intensity plot (d). These intensity profiles were taken from the small rectangular areas marked in (a).

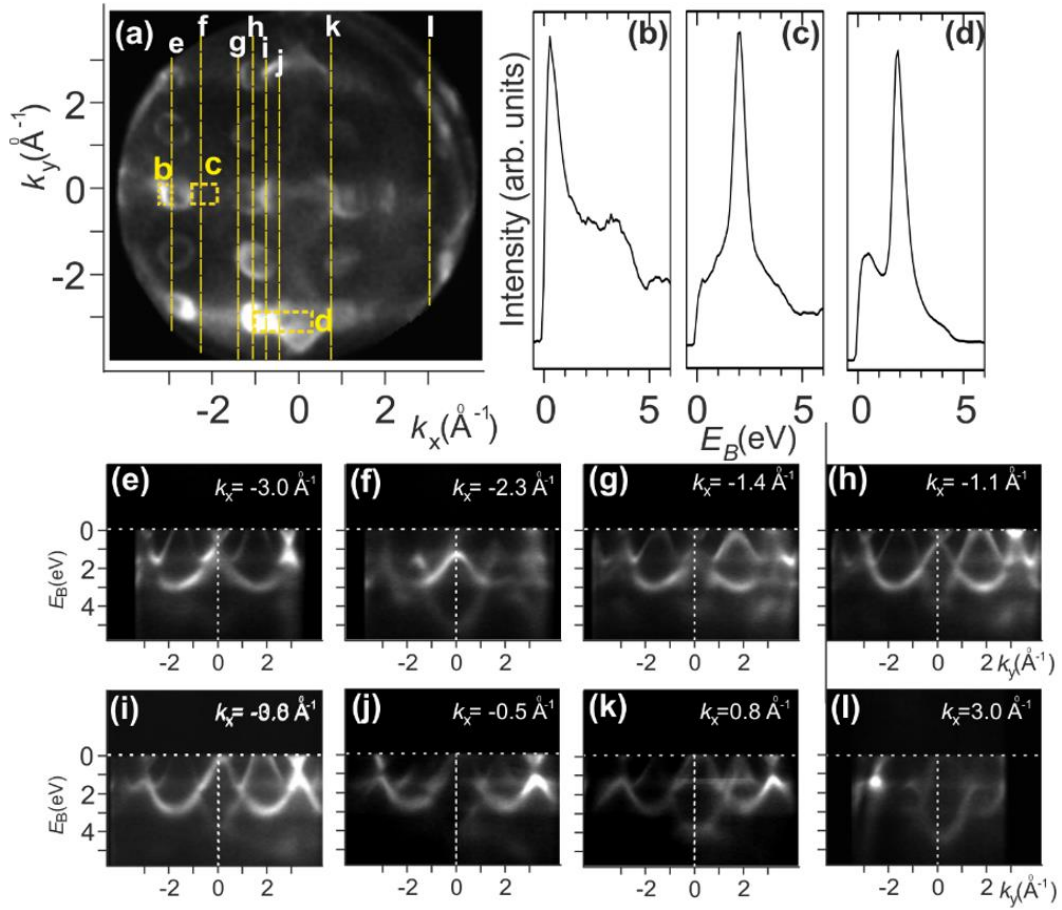

**Figure 6.** Energy- and $k$-dependence of valence-band photoelectron diffraction for Mo(110) at a photon energy of 400 eV. (a) Isosurface at $E_F$; (b-d) energy profiles of the intensities in the areas marked by rectangles in (a). (e-l) cuts along the corresponding dashed lines in (a), revealing several regions of intensity enhancement. Although the photon momentum is rather small at 400 eV, the shift of the final-state sphere causes a striking asymmetry in the patterns.

Sequence (e-l) shows $E_B$-$k_y$ cuts through the measured 3D data array at various values of $k_x$ as given in the panels. This sequence reveals how the position for constructive interference moves through the ($E_B$,**k**) parameter space. Panel (e) shows the locally enhanced intensity in the electron ball in the repeated BZ at the lower left rim of pattern (a). The enhancement shows a rather sharp cutoff at a binding energy of ~2 eV. Panel (f) shows the enhanced maximum of a band that stays well below $E_F$. Sequence (g-j) shows how the interference condition crosses the equivalent k-space object in the BZ at the bottom of panel (a). The enhancement exhibits a characteristic "fine structure". Panel (k) extends the analysis to the other side of the bottom BZ and panel (l) returns to the object of (e), but at opposite $k_x$ and $k_y$.

Owing to the much lower kinetic energy the effect of the MEWDOS in Fig. 6 is weaker than in Fig. 4(d,e), but still visible as horizontal stripe of slightly enhanced intensity at $E_B$~ 1.7 eV. Since this diffuse background has a constant energy dependence, the MEWDOS stripes can be eliminated from the data. This was demonstrated in the HAXPES range at 6 keV photon energy [1].

### *3.3 Valence-band PED at very low photon energies*

We found evidence of VBPED in the VUV spectral range in similar experiments at the 10 m NIM beamline (U125, BESSY II, Berlin) using *p*-polarized light with the *E*-vector oriented at 22° from the surface normal. Fig. 7 shows results for Re(0001), taken at photon energies of 15, 15.5 and 16 eV. An intense spot of constructive interference appears in the centre of the $k_x$-$k_y$ momentum images (a,d,g), i.e. in normal emission. The $E_B$-vs-$k_x$ sections (b,e,h) reveal that the enhancement is restricted to a small energy range of ~200 meV width. When varying the photon energy, this spot shifts in binding energy by the same amount, as revealed by the intensity profiles (c,f,i). Clearly, the constructive interference stays at a fixed final-state energy, here at $E_{final}$= 24 eV (note that $E_{final}$ can be considered as the *inner* kinetic energy, including the inner potential according to Eq. (2)).

The quantitative scheme depicted in Fig. 7(l) shows that this transition leads to the centre of the 2$^{nd}$ repeated BZ along $k_z$, confirming the diffraction condition of Eq. (5). Thus, the experimental result indicates that the forward-scattering mechanism discussed in Fig. 3(a) and observed at 460 eV for Mo(110) (Fig. 4) persists in the VUV range. At such low photon energies, the photon momentum is negligible, making the interpretation of VBPED patterns much easier. However, the final state is no longer free-electron like, but deviations from parabolic dispersion occur. Given the lattice constant of Re along (0001) of 445.6 pm, the reciprocal lattice vector is $G_{0001}$ = 1.41 Å$^{-1}$. The centre of the 2$^{nd}$ BZ is at $k_z$=2$G_{0001}$=2.82 Å$^{-1}$. In the vicinity of $E_F$ there is a total bandgap at the Γ-point, see measured Fermi surface in Fig. 7(k). Hence, the point of constructive interference lies on the $E_B$=1 eV isosurface. In turn, the intense spot in the centre suddenly disappears when the photon energy is reduced to below 15 eV because the diffraction condition leads to the band gap. In (b) and (h), the intensity maximum can be located with a precision of about 200 meV. Then Eq. (2) yields an effective mass of $m_{eff}$ = 1.22 $m_e$ with a precision of <1% (assuming $V_0^*$ = 10 eV). This value for $m_{eff}$ is realistic for final states at such low energies.

To verify that the kinematic diffraction conditions do indeed persist even at these low energies, we made sure that constructive interference also occurs when the final-state sphere reaches the upper boundary of the 2$^{nd}$ BZ (AHL-plane, see Fig. 7(j)). Indeed, we found this point at a photon energy of about 32 eV, corresponding to 2.5 $G_{0001}$=3.52 Å$^{-1}$ (dashed circle in Fig. 7(l); data not shown). Here the VBPED effect is clearly visible, allowing to identify the position of the AHL-plane in the same way as the ΓKM-plane in the data shown in Fig. 7. Quantitatively, the enhancement factor is lower when crossing the BZ boundary in comparison to crossing its centre. The zone boundary is the case of backward scattering (Eq. (6); Fig. 3(b)) requiring the maximum possible momentum transfer of $G_z=2k_f$ which may be less favoured than forward scattering. Further details on the comparison of the band-structure mapping of Re(0001) in the VUV and soft X-ray range will be given in [39]. The independent measurements of the VBPED energies for the two high-symmetry planes ΓKM and AHL (Fig. 7(j)) allows the determination of both unknown quantities $m_{eff}$ and $V_0^*$ in Eq. (2), as will be shown in [37,39].

In fact Re(0001) photoemission at hν=15-16 eV exhibits the strongest intensity enhancement which we found so far. Moreover, it is even more confined in energy than in the 460 eV case for Mo (compare Figs. 4(c) and 7(h)). The photoemission transition starts from an itinerant band state with the sample at 40 K. Hence, this result can only be explained in terms of the momentum-transfer model rather than forward scattering from neighbouring atoms in real space. Unlike the measurements in the soft X-ray range, Fig. 7 shows the complete momentum distribution corresponding to the full polar angular range 0-90°. The VBPED effect represents a significant fraction of the integral photoemission signal (PED-enhanced total photocurrent).

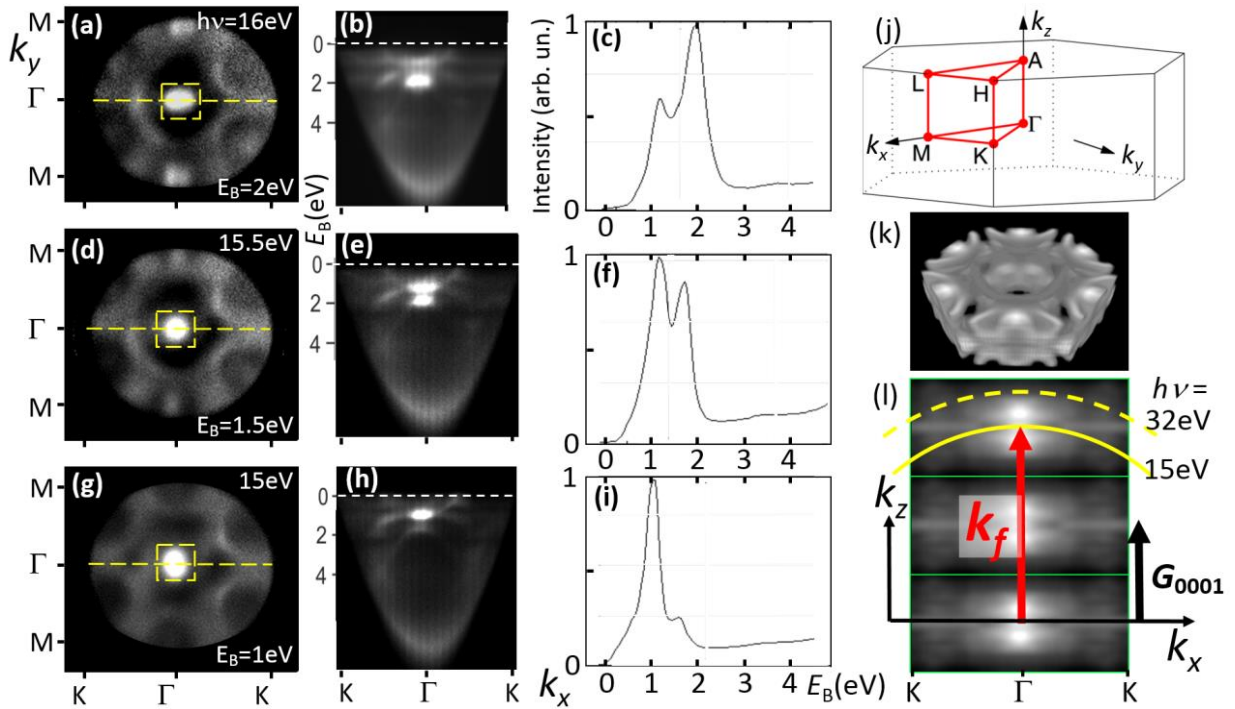

**Figure 7.** Photon-energy dependence of valence-band photoelectron diffraction for Re(0001) between hν=15 and 16 eV. (a,d,g) $k_x$-$k_y$ momentum patterns; (b,e,h) corresponding $E_B$-$k_x$ cuts along the dashed lines in the first column; and (c,f,i) energy profiles of the intensities in the areas marked by rectangles in the first column. The interference condition shifts with photon energy; steps of 0.5 eV correspond to momentum steps of only 0.04 Å$^{-1}$ along $k_z$. The right column shows the BZ (j), the measured Fermi surface (k) and the quantitative transition scheme (l), revealing a transition to the centre of the 2$^{nd}$ repeated BZ. The background pattern is a $k_z$-$k_x$ cut through the measured isosurface of rhenium at $E_B$= 1 eV (panel (k) and background pattern in (l) from [39]).

Despite the surprising agreement with a quasi-free-electron-like final state as evident from Fig. 7, we must keep in mind that three factors make the low-energy case different from the medium- and high-energy cases discussed above: (i) The final state surface in *k*-space will generally deviate from a sphere (i.e. $m_{eff}$ may depend on ***k***). (ii) In addition to the principal final-state band with nearly parabolic dispersion, "secondary" unoccupied bands exist which can hybridize with the principal band and can serve as final states for the photoemission transition. (iii) A significant contribution of multiple scattering is present.

Factor (i) is analogous to the deviation of the noble-metal Fermi surfaces from a sphere. The deviations are significant close to the boundaries of the BZ. In previous 1D spin-resolved studies of $k_z$ dispersion (only normal emission observed) no clear evidence on opening of hybridization gaps at the BZ boundaries has been found, even including spin information [40]. Observing a ***k***-dependence of $m_{eff}$ at low energies is hampered by the small photoemission horizon. At hν = 15 eV (full sphere in Fig. 7(l)) the horizon is only 1.5 Å$^{-1}$ which does not allow to observe the Γ-L or Γ-H directions in k-space, cf. Fig. 7(j). Hence recording the function $m_{eff}$(***k***) would require measurements of several samples with different surface orientations, which is beyond the scope of this paper.

Factor (ii), a flat secondary band acting as final state would violate an important precondition of our model, namely the continuous increase of $k_f$ with increasing energy. The signal from such transitions would mix with the free-electron-like final state transitions in identical energy isospheres. In spin-resolved $k_z$–dependent studies Mueller et al. [40] indeed observed traces of such transitions for Ir(111). However, these signals are very weak and the signals close to the Fermi level (E and F in Fig. 1 of [40]) were most likely masked by the Ir(111) surface state.

There is no doubt that at such low energies factor (iii), i.e. multiple scattering contributes significantly. So additional regions of intensity enhancement will be present, as is well-known from I(V)-measurements in LEED [41]. In a sequence of measurements for Re(0001) at about 40 different photon energies in the range of hν=12-32 eV, covering more than half of the 3D BZ (mostly in photon-energy steps of 0.5 eV), we did not observe further spots of local intensity enhancement. Significant enhancement only occurred for the strong forward scattering when reaching the ΓKM-plane and the somewhat weaker backward scattering when reaching the AHL-plane at the upper border of the 2$^{nd}$ BZ. The small size of the final-state isosphere and its strong curvature does not allow for many conditions of constructive interference. However, the lack of further enhancement spots might also indicate that multiple-scattering effects smear out the sharp kinematic resonances, which we observed at soft X-ray energies.

## 4. Summary and Conclusions

Addressing the cross-section problem in photoemission in the X-ray range, we have developed a new spectroscopic method that combines full-field momentum imaging using a cathode-lens type electron microscope with time-of-flight parallel energy recording. The high parallelization of this approach facilitates recording of a *k*-field of view comprising several Brillouin zones (BZs) and the full *d*-band complex of a transition metal in a single "exposure". Additional variation of the photon energy yields the full information on the electronic valence-band structure in 4-dimensional ($E_B$,**k**) parameter space [8]. Such 4D data arrays taken for a number of transition metals uncover a special manifestation of valence-band photoelectron diffraction (VBPED) that is quite different from previous results in the literature.

The present study elucidates the dependence on the final-state energy $E_{final}$ and momentum vector **k**. The goal was to understand the origin of the appearance of strong intensity enhancements in small regions of ($E_B$,**k**) space ($\Delta k$ few hundredths of Å$^{-1}$, $\Delta E_B$ few hundred meV) and to develop a method for their quantitative analysis. Understanding the origin of VBPED-induced "artefacts" is crucial for quantitative band mapping. Moreover, VBPED allows for a determination of the lattice structure in real space in a similar manner to core-level PED. This opens a path for relating the electronic structure in reciprocal space and lattice structure in real space in a simultaneous measurement.

The first important finding is that *VBPED-induced local intensity enhancement is a very general phenomenon*, as we confirmed for various bcc, hcp and fcc 3d metals at many photon energies. A key result shining light on the mechanism is the *striking asymmetry in the observed patterns, which is induced by the transfer of the photon momentum to the photoelectron*. The effect of the photon momentum on ARPES spectra is well-known and has been observed before [1,2]. However, its asymmetric appearance in *k*-distributions was unexpected. The transfer of photon momentum **k**$_{hv}$ causes a shift of the sphere representing the free-electron-like final states in *k*-space. In the studied photon-energy range of 400 to 1700 eV, this shift increases from 0.20 to 0.86 Å$^{-1}$; the latter value corresponds to half of the BZ diameter.

We introduce a new description of VBPED, where a reciprocal lattice vector **G**$_{hkl}$ is added to the final-state momentum vector, leading to constructive interference. This description is *equivalent to the Laue condition* **k**$_f'$ – **k**$_f$ = **G**$_{hkl}$. The momentum vector of the Bloch wave of the initial state as well as the photon momentum are included in the total momentum balance. VBPED in a given system becomes visible when the final-state sphere in *k*-space (displaced by **k**$_{hv}$) is overlaid with the periodic pattern of initial-state energy isosurfaces. The sphere defines all sets of states to which scattering can take place under the elastic condition. Including the *k*-vector of the itinerant initial state, imposes that a valence band must cross the region in ($E_{final}$,**k**) parameter space where the Laue condition is fulfilled, thus leading to a local intensity modulation of this band feature.

Using a graphical representation resembling the Ewald construction, we were able to quantitatively analyse various cases of VBPED for Mo(110), W(110), Ir(111) and Re(0001). Intensity enhancements by VBPED have been studied in normal and off-normal emission directions. Contributions from low-indexed and high-indexed vectors **G**$_{hkl}$ is evident for Mo(110) at hν= 1700eV; the cases **G**$_{-1\ 1\ 0}$ and **G**$_{-2\ 14\ 0}$ correspond to the zero-order and a higher-order Laue zone. Diagrams like Fig. 5 illustrate and allow us to quantify momentum conservation in *Fermi's Golden Rule* at X-ray energies in an easy and intuitive way. The asymmetry in the results proves that indeed the full final-state sphere is shifted as explained in Fig. 5.

In conclusion, this study reveals that photoelectron diffraction in angular- or momentum-resolved photoemission plays an important role in electronic band mapping, in particular at high energies. The most striking advantage of the proposed model is that it does not require assumptions about spatial localization of the initial state. The only relevant precondition is that the system of isoenergetic surfaces in *k*-space is periodic (demanding a periodic lattice in real space). Application of the Laue condition represents the kinematic approximation, neglecting multiple-scattering processes. This model is a good approximation in the X-ray range, whereas at low energies multiple scattering will lead to additional conditions for constructive interference. Nevertheless we found pronounced VBPED in photoemission from Re(0001) at very low photon energies of 15-16 eV, explained by the kinematic model in forward scattering in normal emission.

Both the VBPED patterns and the broken symmetry due to the transfer of photon momentum show up in the total final state of the photoelectron, thus these phenomena can be considered as matrix element effects. The present results demonstrate that the approximation of a *multiplicative superposition* of band features and VBPED patterns using the Laue conditions in 3D *k*-space captures the complex intensity variations.

Although the present data were obtained for bulk metals, the same description is valid for photoemission of layered materials and thin films. The kinematic model illustrated in Figs. 3(f) and 5(d) applies also for photoemission from an epitaxial layer (e.g. a monolayer) and VBPED involving scattering at the substrate lattice. For such systems momentum-transfer processes are restricted to vectors ***k'*** from the lower hemisphere. The special case of Umklapp processes into final states with $k_\perp = 0$ has been recently discussed in Refs. [30,42].

The present results are particularly important for mapping of the circular and linear dichroism and spin texture in the photoemission patterns, important measurements for spintronic and quantum materials, which are rapidly becoming more possible. VBPED-enhanced local regions bear the danger of artefacts for dichroism and spin distributions. Furthermore, the proposed model is relevant for the emerging field of time-resolved PED. Combining such momentum microscopy with standing-wave excitation in ARPES, with the latter first demonstrated recently [43] can also provide element sensitivity in the emission, thus selecting the VBPED profile appropriate to that atomic type, and providing yet more detail on electronic structure. The range of measurements for which the VBPED effects discussed here are relevant is thus enormous.

## Acknowledgements

Sincere thanks go to K. Rossnagel (U. Kiel, Germany), P. Hofmann (U. Aarhus, Denmark) and C.S. Fadley (U.C. Davis, USA) for a critical reading and valuable comments on our manuscript. Further thanks are due to the staff of beamline P04 (PETRA III, Hamburg), in particular to J. Viefhaus and M. Hoesch, and to the staff of the 10m NIM (BESSY II, Berlin), in particular to P. Baumgärtel, for excellent support during several beamtimes and for fruitful discussions. Last not least we thank A. Oelsner and team (Surface Concept GmbH) for their permanent support of our work. Funding by BMBF (project 05K16UM1) and DFG (projects SCHO 341/16-1 and Transregio SFB TR49) is gratefully acknowledged.

# References

[1] A. X. Gray, C. Papp, S. Ueda, B. Balke, Y. Yamashita, L. Plucinski, J. Minár, J. Braun, E. R. Ylvisaker, C. M. Schneider, W. E. Pickett, H. Ebert, K. Kobayashi and C. S. Fadley, *Probing bulk electronic structure with hard X-ray angle-resolved photoemission,* Nature Mat. **10**, 759 (2011)

[2] L. Plucinski, J. Minár, B. C. Sell, J. Braun, H. Ebert, C. M. Schneider and C. S. Fadley, *Band mapping in higher-energy x-ray photoemission: Phonon effects and comparison to one-step theory,* Phys. Rev. B **78**, 035108 (2008); C. Papp L. Plucinski, J. Minar, J. Braun, H. Ebert, C. M. Schneider and C. S. Fadley *Band mapping in x-ray photoelectron spectroscopy: an experi-mental and theoretical study of W(110) with 1.25 keV excitation*, Phys. Rev. B **84**, 045433 (2011)

[3] C. S. Fadley, *Looking Deeper: Angle-Resolved Photoemission with Soft and Hard X-rays*, Synchrotron Radiation News **25**, 26 (2012)

[4] V. N. Strocov, T. G. E. Cirlin, J. Sadowski, J. Kanski and R. Claessen, *GaSb/GaAs quantum dot systems: in situ synchrotron radiation X-ray photoelectron spectroscopy study*. Nanotechnology **16**, 1326 (2005); Xu, S.-Y., N. Alidoust, I. Belopolski, Z. Yuan, G. Bian, T.-R. Chang, H. Zheng, V. N. Strocov, D. S. Sanchez, G. Chang, C. Zhang, D. Mou, Y. Wu, L. Huang, C.-C. Lee, S.-M. Huang, B. Wang, A. Bansil, H.-T. Jeng, T. Neupert, A. Kaminski, H. Lin, S. Jia and M. Zahid Hasan *Discovery of a Weyl fermion state with Fermi arcs in niobium arsenide*. Nat. Phys. **11**, 748 (2015); B. Q. Lv, N. Xu, H. M. Weng, J. Z. Ma, P. Richard, X. C. Huang, L. X. Zhao, G. F. Chen, C. E. Matt, F. Bisti, V. N. Strocov, J. Mesot, Z. Fang, X. Dai, T. Qian, M. Shi and H. Ding, *Observation of Weyl nodes in TaAs*, Nat. Phys. **11**, 724 (2015)

[5] M. Kobata, S. Fujimori, Y. Takeda, T. Okane, Y. Saitoh, K. Kobayashi, H. Yamagami, A. Nakamura, M. Hedo, T. Nakama and Y. Ōnuki, *Electronic structure of $EuAl_4$ studied by photoelectron spectroscopy*, J. Phys. Soc. Jpn. **85**, 094703 (2016)

[6] L.-P. Oloff, M. Oura, K. Rossnagel, A. Chainani, M. Matsunami, R. Eguchi, T. Kiss, Y. Nakatani, T. Yamaguchi, J. Miyawaki, M. Taguchi, K. Yamagami, T. Togashi, T. Katayama, K. Ogawa, M. Yabashi and T. Ishikawa, *Time-resolved HAXPES at SACLA: probe and pump pulse-induced space-charge effects,* New J. of Phys. **16**, 123045 (2014); L.-P. Oloff, A. Chainani, M. Matsunami, K. Takahashi, T. Togashi, H. Osawa, K. Hanff, A. Quer, R. Matsushita, R. Shiraishi, M. Nagashima, A. Kimura, K. Matsuishi, M. Yabashi, Y. Tanaka, G. Rossi, T. Ishikawa, K. Rossnagel and M. Oura *Time-resolved HAXPES using a microfocused XFEL beam: From vacuum space-charge effects to intrinsic charge-carrier recombination dynamics* Scientific Reports **6**, 35087 (2016)

[7] G. Berner, M. Sing, F. Pfaff, E. Benckiser, M. Wu, G. Christiani, G. Logvenov, H.-U. Habermeier, M. Kobayashi, V. N. Strocov, T. Schmitt, H. Fujiwara, S. Suga, A. Sekiyama, B. Keimer and R. Claessen, *Dimensionality-tuned electronic structure of nickelate superlattices explored by soft-x-ray angle-resolved photoelectron spectroscopy*, Phys. Rev. B **92**, 125130 (2015)

[8] K. Medjanik, O. Fedchenko, S. Chernov, D. Kutnyakhov, M. Ellguth, A. Oelsner, B. Schönhense, T. R. F. Peixoto, P. Lutz, C.-H. Min, F. Reinert, S. Däster, Y. Acremann, J. Viefhaus, W. Wurth, H. J. Elmers and G. Schönhense, *Direct 3D mapping of the Fermi surface and Fermi velocity*, Nature Mat. **16**, 615 (2017)

[9] C. S. Fadley, *Photoelectron Diffraction*, Phys. Scripta **T17**, 39 (1987); C. S. Fadley, *Synchrotron Radiation Research: Advances in Surface Science*, ed. by R. Z. Bachrach, Plenum Press, New York Vol. **2**, 421–518 (1992)

[10] C. S. Fadley, *X-ray photoelectron spectroscopy: Progress and perspectives,* Journal of Electron Spectroscopy and Related Phenomena **178–179,** 2 (2010)

[11] D. P. Woodruff, *Photoelectron diffraction: past, present and future,* J. Electron Spectrosc. Relat. Phenom. **126,** 55–65 (2002); D. P. Woodruff, *Surface structural information from photo-electron diffraction*, J. Electron Spectrosc. Relat. Phenom. **178-179**, 186 (2010)

[12] M. V. Kuznetsov, I. I. Ogorodnikov, D. Yu. Usachov, C. Laubschat, D. V. Vyalikh, F. Matsui and L. V. Yashina, *Photoelectron Diffraction and Holography Studies of 2D Materials and Interfaces*, J. Phys. Soc. Japan **87**, 061005 (2018)

[13] Ch. Sondergaard, Ch. Schultz, M. Schonning, S. Lizzit, A. Baraldi, S. Agergaard, H. Li and Ph. Hofmann, *Symmetry-resolved density of states from valence band photoelectron diffraction*, Phys. Rev. B **64**, 245110 (2001)

[14] J. Osterwalder *Photoelectron Spectroscopy and Diffraction,* in 'Handbook on Surface and Interface Science', ed. by K. Wandelt, Wiley-VCH, Weinheim, Vol. **1**, pp. 151-214 (2011)

[15] A. Winkelmann, C. S. Fadley and F. J. Garcia de Abajo, *High-energy photoelectron diffraction: model calculations and future possibilities*, New J. of Phys. **10,** 113002 (2008)

[16] P. Krüger, F. Da Pieve and J. Osterwalder, *Real-space multiple scattering method for angle-resolved photoemission and valence-band photoelectron diffraction and its application to Cu(111)*, Phys. Rev. B **83**, 115437 (2011)

[17] J. Osterwalder, T. Greber, S. Hüfner and L. Schlapbach, *X-ray photoelectron diffraction from a free-electron-metal valence band: Evidence for hole-state localization,* Phys. Rev. Lett. **64**, 2683 (1990); J. Osterwalder, T. Greber, P. Aebi, R. Fasel and L. Schlapbach, *Final-state scattering in angle-resolved ultraviolet photoemission from copper*, Phys. Rev. B **53**, 10209 (1996)

[18] G. S. Herman, T. T. Tran, K. Higashiyama and C. S. Fadley, *Valence Photoelectron Diffraction and Direct Transition Effects*, Phys. Rev. Lett. **68**, 1204 (1992)

[19] B. Sinkovic, B. Hermsmeier and C. S. Fadley, *Observation of Spin-Polarized Photoelectron Diffraction,* Phys. Rev. Lett. **55**, 1227 (1985)

[20] B. Hermsmeier, J. Osterwalder, D. J. Friedmann and C. S. Fadley, *Evidence for a High-Temperature Short-Range-Magnetic-Order Transition in MnO(001),* Phys. Rev. Lett. **62**, 478 (1989)

[21] H. Daimon, T. Nakatani, S. Imada. S. Suga, Y. Kagoshima and T. Miyahara, *Strong Circular Dichroism in Photoelectron Diffraction from Nonchiral, Nonmagnetic Material–Direct Observation of Rotational Motion of Electrons,* Jpn. J. Appl. Phys. **32,** L 1480 (1993); H. Daimon, T. Nakatani, S. Imada and S. Suga, *Circular Dichroism from Non-Chiral and Non-Magnetic Materials Observed with Display-type Mirror Analyzer*, J. Electron Spectrosc. Relat. Phenom., **76**, 55 (1995); H. Daimon et al., Surface Science **408**, 260-267 (1998)

[22] S. Kenji, N. Maejima, H. Nishikawa, T. Matsushita and F. Matsui, *Development of Micro-Photoelectron Diffraction at SPring-8 BL25SU*, Surf. Sci. Nanotech. **14**, 59 (2016)

[23] K. Siegbahn, U. Gelius, H. Siegbahn and E. Olson, *Angular Distribution of Electrons in ESCA Spectra from a Single Crystal,* Phys. Scripta **1**, 272 (1970); C. S. Fadley and S. A. L. Bergstrom, *Angular Distributions of Photoelectrons from a Metal Single Crystal,* Phys. Letters 35A, 375 (1971)

[24] M. von Laue, *Materiewellen und ihre Interferenzen*, Akademische Verlagsgesellschaft Geest & Portig, Leipzig (1948)

[25] S. M. Goldberg, R. J. Baird, S. Kono, N. F. T. Hall and C. S. Fadley, *Explanation of XPS Core-Level Angular Distributions for Single-Crystal Copper in Terms of Two-Beam Kikuchi-Band Theory*, J. Electron Spectrosc. **21**, 1 (1980)

[26] R. Trehan, C. S. Fadley and J. Osterwalder, *Single-Scattering Cluster Description of Substrate X-ray Photoelectron Diffraction and Its Relationship to Kikuchi Bands,* J. Electron Spectrosc. **42**, 187 (1987)

[27] F.J. Garcia de Abajo, M.A. Van Hove, C.S. Fadley, *Multiple scattering of electrons in solids and molecules: A cluster-model approach*, Phys. Rev B **63**, 075404 (2001); and C. S. Fadley, private communication (2018)

[28] J. Braun, J. Minar, S. Mankovsky, L. Plucinski, V. N. Strocov, N. B. Brookes, C. M. Schneider, C. S. Fadley and H. Ebert, *Exploring the XPS-limit in hard x-ray angle-resolved photoemission spectroscopy by fully temperature-dependent one-step theory,* Phys. Rev. B **88**, 205409 (2013)

[29] A. Winkelmann, A. A. Ünal, C. Tusche, M. Ellguth, C.-T. Chiang and J. Kirschner, *Direct k-space imaging of Mahan cones at clean and Bi-covered Cu(111) surfaces*, New J. of Physics **14**, 083027 (2012)

[30] A. Winkelmann, M. Ellguth, C. Tusche, A. A. Ünal, J. Henk and J. Kirschner, *Momentum-resolved photoelectron interference in crystal surface barrier scattering,* Phys. Rev. B **86**, 085427 (2012)

[31] J. Minar, J. Braun and H. Ebert, *Recent developments in the theory of HARPES* J. Electron Spectrosc. Relat. Phenom. **190**, 159 (2013)

[32] E. W. Plummer and W. Eberhardt, *Angle-resolved photoemission as a tool for the study of surfaces*, Advance in Chemical Physics, Ed. by I. Prigogine, S. A. Rice, John Wiley &, Vol. XLIX, 533-656 (1982)

[33] S. Hüfner, *Photoelectron Spectroscopy – Principles and Applications*, Springer Berlin, (2003)

[34] E. Bauer, *The possibilities for analytical methods in photoemission and low-energy electron microscopy,* Ultramicroscopy **36,** 52-62 (1991); L. H. Veneklasen, *Design of a spectroscopic low-energy electron microscope,* Ultramicroscopy **36,** 76-90 (1991)

[35] G. Schönhense, K. Medjanik, S. Chernov, D. Kutnyakhov, O. Fedchenko, M. Ellguth, D. Vasilyev, A. Zaporozhchenko, D. Panzer, A. Oelsner, C. Tusche, B. Schönhense, J. Braun, J. Minár, H. Ebert, J. Viefhaus, W. Wurth and H. J. Elmers, *Spin-Filtered Time-of-Flight k-Space Microscopy of Ir – Towards the "Complete" Photoemission Experiment*, Ultramicroscopy **183**, 19–29 (2017)

[36] O. Fedchenko, K. Medjanik, S. Chernov, D. Kutnyakhov, M. Ellguth, A. Oelsner, B. Schön-hense, T. R. F. Peixoto, P. Lutz, C.-H. Min, F. Reinert, S. Däster, Y. Acremann, J. Viefhaus, W. Wurth, J. Braun, J. Minár, H. Ebert, H. J. Elmers and G. Schönhense, *4D Texture of Circular Dichroism in Soft-X-Ray Photoemission From a Non-Magnetic Solid,* Submitted (2018)

[37] S. Babenkov, K. Medjanik, D. Vasilyev, M. Ellguth, A. Zymakova, C. Tusche, A. Quer, F. Diekmann, S. Rohlf, M. Kalläne, K. Rossnagel, Y. Acremann, D. Kutnyakhov, W. Wurth, J. Viefhaus, G. Schönhense and H.-J. Elmers, *Bulk-Sensitive Electronic-Structure mapping of Mo in Soft and Tender X-ray Range*, in preparation

[38] M. P. Seah and W. A. Dench, *Quantitative Electron Spectroscopy of Surfaces: A Standard Data Base for Electron Inelastic Mean Free Paths in Solids*, Surface and Interface Analysis **1**, 2 (1979)

[39] K. Medjanik, D. Vasilyev, S.Babenkov, M. Ellguth, B. Schönhense, J Viefhaus, H.-J. Elmers and G. Schönhense , *Fermi Surface, Fermi-Velocity and Circular Dichroism of Rhenium,* in preparation

[40] N. Müller, B. Kessler, B. Schmiedeskamp, G. Schönhense, U. Heinzmann, *Spin-resolved photoemission from Ir(111): Transitions into a secondary band and energetic position of the final state bands*, Solid State Commun. **61**, 187 (1987); A. Eyers, F. Schäfers, G. Schönhense, U. Heinzmann, H. P. Oepen, K. Hünlein, J. Kirschner, G. Borstel, *Characterization of symmetry properties*

*of Pt(111) electron bands by means of angle-, energy-, and spin-resolved photoemission with circularly polarized synchrotron radiation*, Phys. Rev. Lett. **52**, 1559 (1984)

[41] R.S. Zimmer and W.D. Robertson, *LEED from Re(0001); and in comparison from Be(0001)*, Surf. Sci. **43**, 61 (1974).

[42] A. Zaporozhchenko-Zymaková, D. Kutnyakhov, K. Medjanik, C. Tusche, O. Fedchenko, S. Chernov, M. Ellguth, S.A. Nepijko, H.J. Elmers and G. Schönhense, *Momentum-resolved photoelectron absorption in surface barrier scattering on Ir(111) and graphene/Ir(111),* Phys. Rev. B **96**, 155108 (2017)

[43] S. Nemšák, S. Döring, C. Schlüter, M. Eschbach, E. Mlynczak, T.-L. Lee, L. Plucinski, J. Minar, J. Braun, H. Ebert, C. M. Schneider, C.S. Fadley, *Hard x-ray standing-wave angle-resolved photoemission: element- and momentum-resolved band structure for a dilute magnetic semiconductor*, Nature Communications, in print: https://arxiv.org/abs/1801.06587